\documentclass[letterpaper, 10 pt, conference]{IEEEtran}  

\usepackage{times}
\usepackage{verbatim}
\usepackage{cite}
\usepackage{graphicx}
\usepackage{color}
\usepackage{authblk}
\usepackage{commath}
\usepackage{tabulary}
\usepackage{caption}
\usepackage{xr}

\title{
Two stage cluster for resource optimization with Apache Mesos
}

\author[1]{Gourav Rattihalli}
\author[1]{Pankaj Saha}
\author[1]{Madhusudhan Govindaraju}
\author[2]{Devesh Tiwari}
\affil[1]{{\it{Cloud and Big Data Lab, State University of New York (SUNY) at Binghamton}}}
\affil[2]{\it{Northeastern University}}
\affil[ ]{\textit {\{grattih1, psaha4, mgovinda\}@binghamton.edu and d.tiwari@northeastern.edu}}

\begin{document}
\date{}
\maketitle
\thispagestyle{empty}
\pagestyle{empty}

\begin{abstract}

As resource estimation for jobs is difficult, users
often overestimate their requirements. Both commercial clouds and academic campus clusters suffer from low resource utilization and long wait times as the resource estimates for jobs,
provided by users, is inaccurate. We present an approach to statistically estimate the actual resource requirement of a job in
a ”Little” cluster before the run in a ”Big” cluster. The initial estimation on the little cluster gives us a view of how much actual resources a job requires. This initial estimate allows us to accurately allocate  resources for the pending jobs in  the queue and thereby improve throughput and resource utilization. In our experiments, we determined resource utilization estimates with an average accuracy of 90\% for memory and 94\% for CPU, while we make better utilization of memory by an average of 22\% and CPU by 53\%, compared to the default job submission methods on Apache Aurora and Apache Mesos.

\end{abstract}
%



\section{Introduction}

Scientific jobs can vary vastly in the resources they consume
depending on the core application kernel, input size, and other
application specific parameters. In the emerging cloud
infrastructures, users are allocated a cluster of VMs or bare metals,
and the utilization of the allocated cluster depends on the accuracy
of user's estimate for resources (CPU, Memory, Disk, Network
bandwidth, I/O) required for each job. Inaccurate user estimates,
which directly causes under-utilization of resources, also affects
overall throughput, increases wait times for jobs in the queue, and
costs higher as VMs are priced based on their configurations.

Huge wait times is a well-documented problem both for HPC clusters in
academia, and cloud infrastructures used by the industry and
scientific researchers. For example, Marathe
et. al~\cite{Marathe2013ACloud} ran experiments on an Amazon EC2
cluster and ``measured wait times of 146 seconds, 189 seconds and 244
seconds to acquire 16, 32 and 64 EC2 nodes." They further observed
that wait times at a Lawrence Livermore National Laboratory (LLNL)
cluster was orders of magnitude higher than at Amazon EC2.

In the analysis done by Reiss et. al. using the Google trace, which
was collected for a 12000 node cluster, it was found that the CPU
allocation for high priority production jobs was at an average of
70\%, whereas the CPU usage stood at an average of just
25\%. Similarly, for memory utilization the allocations were at an
average of 60-70\% but the utilization was at an average of just
30-35\%~\cite{Reiss2012HeterogeneityScale}.

At Twitter, their large-scale cluster uses Mesos for resource
management. This cluster with thousands of server class nodes has
reservations reaching 80\% but the utilization has been found be to
consistently below 20\%~\cite{DelimitrouQuasar:Management}.

At SUNY Binghamton's Spiedie campus cluster, the snapshot of a single day's data shows that overall users requested significantly more CPU resources than required - the number of cores requested was 7975 but the actual usage was 4415 cores.

When commercial cloud facilities such as Amazon EC2 and Microsoft Azure operate at very low resource utilization, it increases the cost
to operate the facilities, and thereby the cost for end users. The
usage of schedulers and cloud management tools such as Apache
Mesos~\cite{Hindman2011Mesos:Center} mitigates this to some extent as they facilitate co-scheduling of tasks on the same node.  However, the advantages of Mesos are limited when users request and reserve more resources than required by their applications. It has been observed that in 70\% of the cases, the users request 
significantly more than the required amount of resources~\cite{DelimitrouQuasar:Management}. It is to be noted that under-reserving the resources is also a significant hurdle. Mesos co-schedules jobs on the shared infrastructure, isolates them using containers/cgroups, but kills the jobs that attempt to exceed their reserved resources.

Open Source Big Data and cloud technologies such as Mesos
\cite{Hindman2011Mesos:Center}, Omega\cite{Schwarzkopf2013Omega},
Yarn\cite{Vavilapalli2013ApacheYARN}, and Torque\cite{TORQUEManager}
require user input to make resource allocation. These allocation
requests are processed by the cloud management software, based on the
chosen policy or default scheduling algorithm, for the placement of
jobs on the cluster. For example, Mesos receives the resource
allocation requests and based on availability and the dominant
resource fairness (DRF) algorithm~\cite{Ghodsi2011DominantTypes},
makes resource offers to frameworks such as Apache
Aurora\cite{ApacheJobs.} and Marathon\cite{Marathon:DC/OS}. These
frameworks can then accept the offers, or reject the offers and
continue the negotiations. Mesos provides a containerized
environment for each application and ensures that resource usage does not
exceed the allocation. Our focus in this paper is to address the resource
estimation problem in the context of jobs submitted to Apache Mesos
via Apache Aurora.

We use an Apache Mesos based cluster to run Docker container based
jobs using the Apache Aurora framework. 

We run each job in a little cluster, for a few seconds, to get a statistical estimate of the resource requirements and
then launch it on the production cluster using that resource
estimate. We describe how this approach can be designed. We consider
the design space to quantify the ratio of the little and big clusters and the policy for packing the jobs in the little and big clusters. We
quantify the gains in terms of CPU utilization, memory utilization,
and overall throughput.

In this paper, we make the following contributions:

\begin{itemize}

\item We present a design to dynamically profile the incoming
jobs in a Mesos-Aurora cluster with the aim of improving the utilization of cluster resources.

\item We propose a two-stage cluster setup - the first stage to
get a quick estimate of the resource requirement and then the second stage to run the application.

\item We present results from a queue of Dockerized scientific
applications, identify the ideal ratio of the size of the two
clusters, and quantify the improvements in resource utilization.

\end{itemize}

In section II, we discuss the problem definition and summarize the technologies used for the experiments. In Section III, we discuss the approach used to solve the defined problem and we also identify the opportunities for optimization with our current setup. In Section IV and V, we discuss the accuracy of our optimization approach and also the limitation of the approach. In Section VI, we discuss the experimental setup and in Section VII we analyze the results of our experimentation. In Section VIII, we present the related work. Section IX discusses the future work and in Section X we present the conclusions.

\section{Problem Definition and Background}
In this section, we define the research problem and summarize the technologies we have used in this project.

\externaldocument{solution}
\subsection{Problem Definition}
CPU, Memory, Disk space, and I/O bandwidth are fundamental resources requested by an application and these are used by Mesos to make a scheduling decision~\cite{Greenberg2015BuildingMesos}. Additionally, Mesos allows an application to specify the resource units in fractions - 0.5 CPU, for example. Mesos allows co-locating multiple applications on the same node and enforces the default, or user-defined policies, on resource sharing. However, Mesos depends on the user specification on how much resources an application needs. Effective resource utilization, cost of acquiring nodes from a cloud or HPC cluster, and throughput of applications in the queue depends significantly on user specified resource estimates. Specifying an accurate resource requirement is critical: requesting fewer resources will get the application killed by Mesos when it tries to exceed allocation, and overestimation, which is the widely prevailing case, hurts utilization, throughput, and cost.

Once VMs are allocated on the cloud, we divide them into a little and big cluster. We use the little cluster to run the job just long enough to get a statistical estimate of the resource requirement for the job and then subsequently launch it again on the big cluster. It is to be noted that Mesos currently does not support job migration, though it is a planned feature for future releases. Our research problem's scope encompasses (1) identifying the ideal ratio of the size of the little and big cluster, (2) the effect of co-allocation of jobs on the little cluster, and (3) the mapping of resource consumption on the little cluster with the usage on the big cluster for the policies of Aurora and Mesos in use for job scheduling.

\subsection{Apache Mesos} The Berkley AMP lab in 2011 released Mesos as a platform for sharing a cluster between many disparate application like Hadoop and Spark~\cite{Greenberg2015BuildingMesos}. Next, Twitter developed it further to make it production-ready to manage a large number of nodes in their data centers. Today, Mesos has grown from a research project to core infrastructure that powers tens of thousands of servers at various companies. Major companies that use Mesos are Twitter, Apple, Cisco, Netflix, Ericsson, Uber, Verizon Labs and eBay~\cite{ApacheMesos}.

Mesos directly supports the use of container technologies like Linux Containers(LXC) and Docker. Therefore, Mesos is an Operating System as well as the Infrastructure Management layer that allows various cluster computing frameworks to efficiently share compute resources\cite{Hindman2011Mesos:Center}.

    \subsection{Apache Aurora}
    Apache Aurora is a framework that is designed to submit, execute and manage jobs to a Mesos cluster. 
Aurora uses the information provided by the Mesos cluster manager to make its scheduling decisions. It has its own domain specific language to configure jobs. Once the job is launched on the cluster, Aurora also manages the job, i.e. if the job experiences failure it reschedules the job on another healthy node.

    \subsection{Chameleon Cloud}
Chameleon Cloud is an infrastructure funded by the National Science Foundation (NSF) under the NSFCloud initiative\cite{Mambretti2015NextSDN}. It consists of 18000 processor cores and over 500 cloud nodes and about 5 petabytes of disk storage. Researchers can use this infrastructure by using pre-configured system configurations or create their own VM images.
 Chameleon cloud also supports ``bare-metal access" as an alternative for researchers who require more control and isolation to their setup. Allocations to users are done in terms of Services Units (SUs), which is a unit that collectively measures the number of VMs, number of cores, total memory, and disk capacity assigned to a user. Typical startup allocations are 20,000 SUs.
    
    \subsection{Docker}
Docker is a lightweight Virtual Machine (VM). It is a container based technology that shares the Operating System with the host\cite{JamesTurnbull2014TheVirtualization}. The Virtual Machine provides an abstraction at the hardware level, while the container virtualizes the system calls. This allows multiple isolated instances of Docker containers to be started on the same host\cite{Adufu2015IsApplications}. {\it cgroups} plays an important role in enforcing allocation of resources such as CPU, Memory and I/O to containers\cite{MarekGoldmann2014ResourceDocker}.
    
    \subsection{Docker Swarm}
Docker Swarm is a native clustering mechanism for Docker. 
 We use the standalone version of Docker Swarm. The standalone version provides a centralized method for collecting resource utilization of every container.
 
    \subsection{Performance Co-Pilot}
Performance Co-Pilot is a lightweight application to collect performance metrics of systems. It was released in 1995 and was originally created by SGI exclusively for its customers. In 2000, it was re-released as free software under GNU LGPL. It can collect metrics from multiple hosts and different Operating Systems\cite{KenMcDonellPerformanceCo-Pilot}. We use Performance Co-Pilot to monitor all the agents and collect metrics to perform the analysis.
\section{Approach And Optimization Space}
A well-known approach to tackle over-allocation is to profile the application before it is executed. There are two ways to profile an application, Static or Off-line profiling and Dynamic or On-line profiling. Let's consider static or off-line profiling - this approach provides a more accurate solution to the problem of over-allocation. By statically profiling the application we can determine what resource an application needs to execute efficiently. However, this approach is time-consuming, as applications can run for a long time and HPC and cloud resources may not be available for multiple runs. Additionally, if the acquired cluster is heterogeneous, which is often the case in cloud infrastructures, then the application would have to be profiled on all the different types of systems. Instead, an effective approach is to dynamically identify the resources required by the application. While this approach is quicker, it can suffer errors in estimation. So, if it were possible to make this approach more accurate, then it would provide a balance between accuracy and efficiency. In this paper, we explore the possibility to optimize the approach of dynamically profiling applications and its effect on performance.

\begin{figure}[h!]
			\includegraphics[width=0.5\textwidth, height=4.5cm]{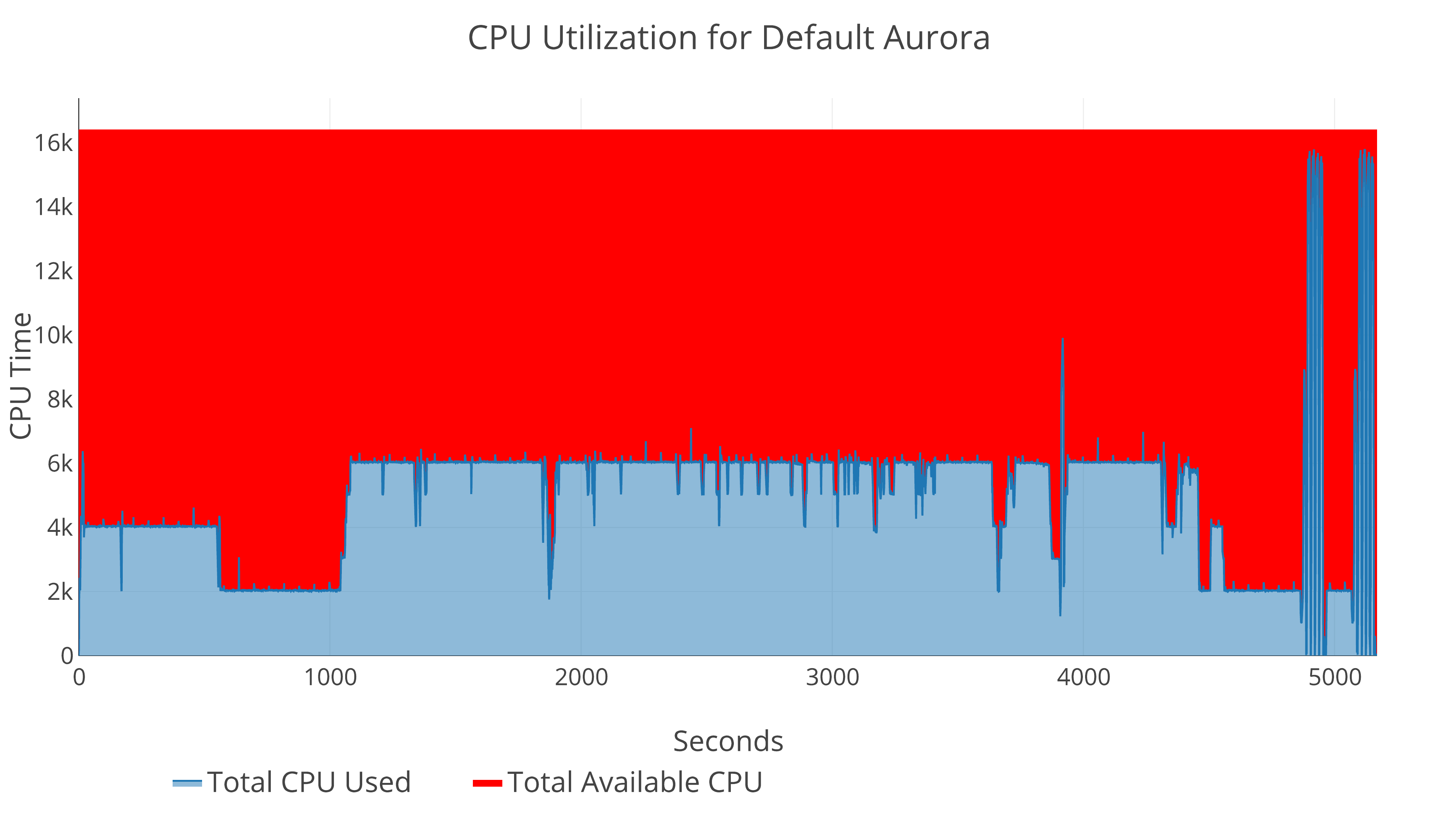}
			\caption{{\it Optimization Space - CPU utilization. The region in red is the unused CPU time.}}
            \vspace{-1.5em}
            \label{fig:optSpaceCpu}
\end{figure}

\begin{figure}[h!]
			\includegraphics[width=0.5\textwidth, height=4.5cm]{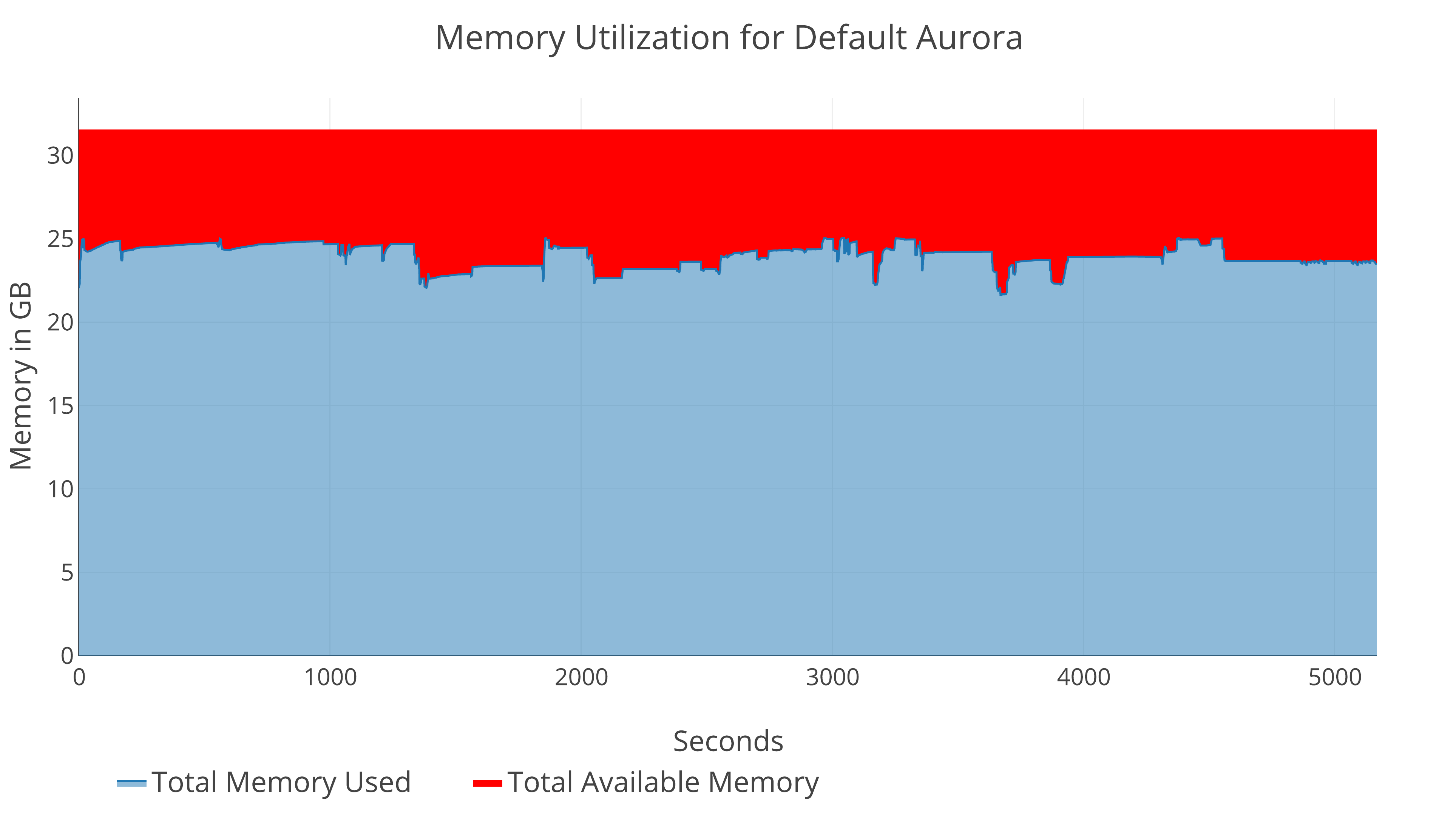}
			\caption{{\it Optimization Space - Memory utilization. The region in red is the unused memory.}}
            \vspace{-1.5em}
            \label{fig:optSpaceMem}
\end{figure}

In the Figure \ref{fig:optSpaceCpu}  and \ref{fig:optSpaceMem} we present the case for the need of optimization. The areas in red are the unused resources in terms of memory and CPU time.

\section*{APPROACH}
\subsection{Exclusive Access Optimization Method}
\begin{figure}[h!]
			\includegraphics[width=0.5\textwidth, height=3cm]{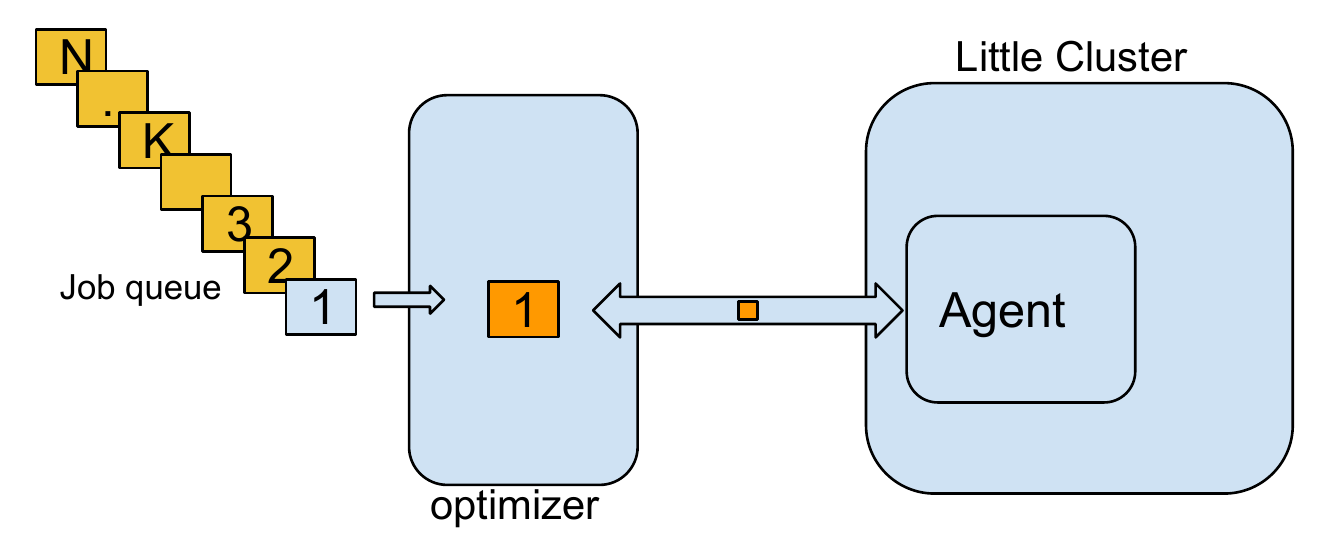}
			\caption{{\it Exclusive Access Optimization Approach. Each application is separately profiled before being launched on the big cluster.}}
            \label{fig:queued}
		\end{figure}

In this approach, we execute the first application on the little cluster and profile it by giving it exclusive access. While an application is being profiled, the other benchmarks wait in the queue. We gather the resource usage information for each application and estimate the optimal resource requirements using the following steps:
\begin{itemize}
\item Record the last five observations of memory usage and CPU usage.
\item If the majority of observations are within the threshold i.e. 95\% C.I. (confidence interval), then calculate the buffer resources. Else, record the next 5 observations of resources. Continue this until the observations are under the C.I.  
\item To calculate the buffer, get the modulus of Standard deviation/positive deviation of the observations.
\[\textit{buffer} = \abs{\sqrt{\frac{1}{N-1} \sum_{i=1}^N (x_i - \overline{x})^2}}\]
Here, x is the observation and N is the number of observations.
\item The optimal resource is the sum of the median of observations and the buffer.
\[\textit{Optimal Resource} = {\textit{Median Of Observations} + \textit{buffer}}\]
\item We calculate the buffer so that the application has headroom in terms of available resources. If we do not allocate the buffer, there might be cases where the application may fail/terminate.
\end{itemize}

Once the optimal amount of resources are calculated, the optimizer prepares the configuration required to launch the job on the big cluster.

\subsection{Co-Scheduled Optimization Method}

\begin{figure}[h!]
			\includegraphics[width=0.5\textwidth, height=3cm]{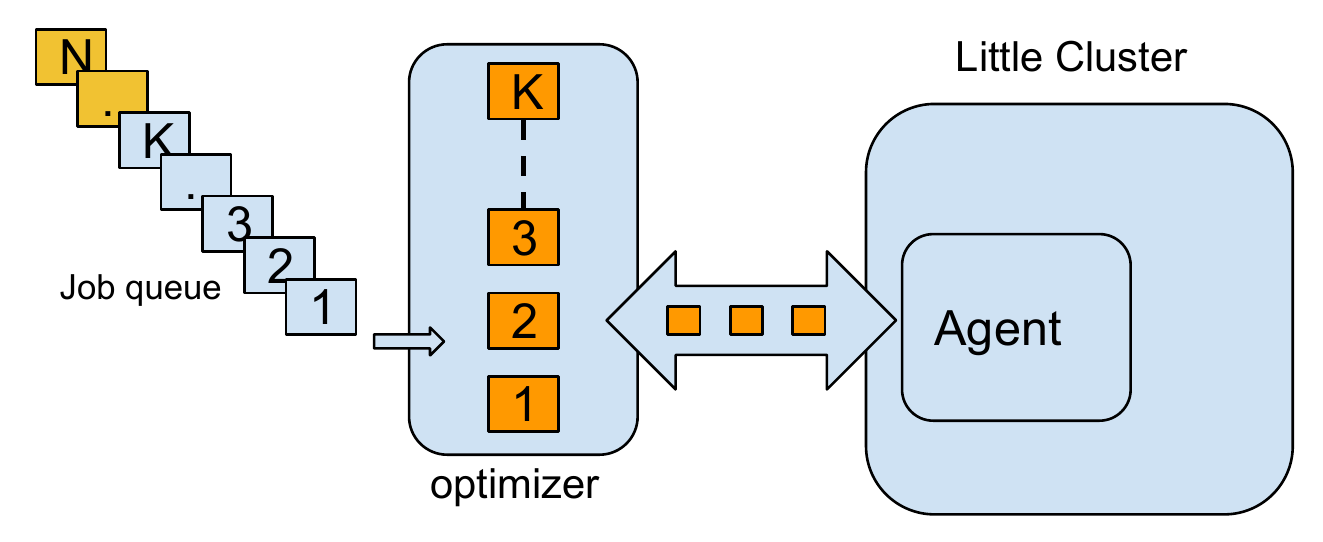}
			\caption{{\it Co-Scheduled Optimization Approach. Applications are co-scheduled on the little cluster and then profiled.}}
            \label{fig:simultaneous}
		\end{figure}
In this approach, we co-schedule applications on the little cluster. The maximum number of applications that can be run on the little cluster is determined by the available resources and the resource requests of the application. This method gives us an estimate of how much resources each application requires when the {\it cgroups} are shared between multiple applications, unlike the previous case wherein each application had 100\% share of the cgroups. This method forces the application to use limited resources. Our optimizer gathers the resource usage information and estimates the optimal resource requirements in the same way as the Exclusive Access approach. The optimizer then creates the required Aurora configuration file, with the updated resource information, to run the application on the big cluster.

\section{EXPERIMENTAL SETUP}
	\subsection{Big-Little setup}
    	\begin{figure}[h!]
			\includegraphics[width=0.5\textwidth, height=4.5cm]{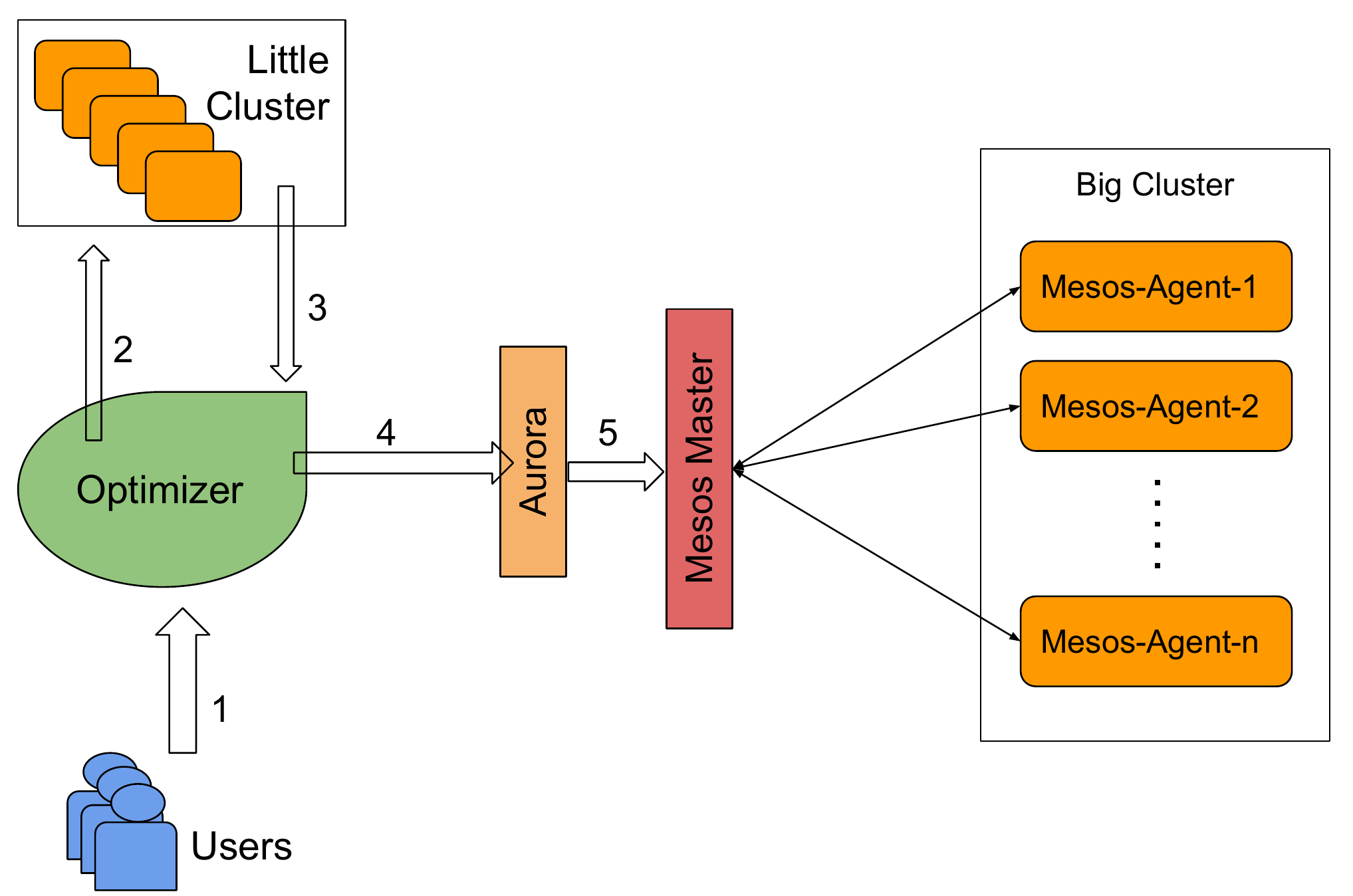}
			\caption{\it{Big-Little Setup with Optimizer}}
            \vspace{-1em}
            \label{fig:Big-Little}
		\end{figure}
        The important components of the experimental setup are the optimizer, big and little clusters. The optimizer is the module that executes the container on the little cluster, in which the nodes are managed by Docker Swarm and collects the resource usage which is then used for the optimized estimate. When a job is submitted to the optimizer, it executes this job on the little cluster for a short duration and collects the resource usage. Once the optimal resources are calculated, it creates the Aurora configuration for the job with the optimized resources and then submits to the Aurora scheduler. Aurora communicates with the Mesos master to schedule the job on the big cluster.
        
Table \ref{desc} provides the information about the equipment, Operating System and other software used in the experimental setup.
\begin{table}[h!]
\small
\centering
\renewcommand{\arraystretch}{1.5}
\begin{tabular}{|l|p{0.55\linewidth}|}
 \hline
 {\bf Equipment/OS/Software} & {\bf Description/Version}\\
 \hline
 Node (13) & Intel Xeon E312xx, 8 core processor at 2.3GHz, 16 GB DDR3 RAM  \\ 
  \hline
  Operating System & Ubuntu 16.04 \\
  \hline
  Docker & 17.06.1-ce \\
  \hline
  Apache Mesos & 1.1.0 \\
  \hline
  Apache Aurora & 0.17.0 \\
  \hline
  \end{tabular}
\captionsetup[table]{skip=6pt}
\caption{\it{Description of the nodes, OS, and software versions used in the experiments.}}
\vspace{-1.5em}
\label{desc}
\end{table}      

    \subsection{Benchmarks/Workloads}
		\subsubsection{Princeton Application Repository for Shared-Memory Computers (PARSEC)} It is a benchmark suite consisting of multithreaded applications. We used PARSEC 3.0. These benchmarks include a diverse set of applications ranging from Financial Analysis to Media Processing\cite{Bienia2008TheImplications} \cite{BieniaPARSECChip-Multiprocessors}\cite{BieniaThe2.0}\cite{BieniaChristianandLi2011BENCHMARKINGMULTIPROCESSORS}.
        
\begin{table}[h!]
\small
\centering
\renewcommand{\arraystretch}{1.5}
\begin{tabular}{|l|p{0.65\linewidth}|}
 \hline
 {\bf Workload} & {\bf Description}\\
 \hline
 Blackscholes & Computational financial analysis application\\ 
  \hline
   Bodytrack & Computer vision application\\  
  \hline
  Canneal & Engineering application \\
  \hline
  Ferret & Similarity search application \\
  \hline
  Fluidanimate & Application consists of animation tasks \\
  \hline
  Freqmine & Data mining application \\
  \hline
  Swaptions & Financial Analysis application \\
  \hline
  Streamcluster & Data mining application \\
  \hline
  DGEMM & Dense-matrix multiply benchmark \\
 \hline
\end{tabular}
\captionsetup[table]{skip=6pt}
\caption{\it{Description of the benchmarks from PARSEC used in the experiments}}
\vspace{-1.5em}
\end{table}    
    
 \subsection{Setup Ratios}
 We setup experiments to determine the ideal best ratio of the size of the little and big cluster. We ran experiments with a queue of 90 applications that had a mix of CPU and memory intensive resource requirements. The different setups of the big and little clusters had the ratios ranging from 1:2 to 1:12. This means for every 12 machines in the big cluster there was one machine in the little cluster. 
\section{Accuracy of our Dynamic Resource Estimation}
In Table \ref{table:memoryaccuracy} and \ref{table:cpuaccuracy}, we compare and analyze our optimization approaches with respect to memory and CPU respectively. The full runs are done with static profiling, which means the benchmarks were profiled from start to finish. The partial runs are profiled with the proposed optimizing system. As we can see from Table \ref{table:memoryaccuracy} the error percentages are under 10\% with the exception of {\it Canneal}, {\it Ferret} and {\it Swaptions} for memory.  As for CPU, the estimates are better, with the exception of {\it Bodytrack} and {\it DGEMM}.

\begin{table}[h!]
\centering
\small
\renewcommand{\arraystretch}{1.5}
\begin{tabular}{ | c | c | c | c | }
 \hline
 {\bf Workload} & {\bf Full Run} & {\bf Partial Run} & {\bf Error} \\
 \hline
 Blackscholes & 1234.31 & 1222.60 & 0.96\% \\ 
 Bodytrack & 970.14 & 1077.65 & 9.98\% \\  
 Canneal & 966.60 & 875.71 & 10.38\% \\
 Ferret & 212.03 & 284.96 & 25.59\% \\
 Fluidanimate & 541.2 & 541.0 & 0.04\% \\
 Freqmine & 825.01 & 794.86 & 3.79\% \\
 Streamcluster & 106.96 & 107.66 & 0.65\% \\
 Swaptions & 4.56 & 3.188 & 43.03\% \\
 DGEMM & 28.4 & 26.41 & 7.54\% \\
 \hline
\end{tabular}
\captionsetup[table]{skip=6pt}
\caption{\it{Memory usage comparison of Static and Dynamic Profiling for well known application workloads. (Units: MB)}}
\vspace{-0.9em}
\label{table:memoryaccuracy}
\end{table}

\begin{table}[h!]
\small
\centering
\renewcommand{\arraystretch}{1.5}
\begin{tabular}{ | c | c | c | c | }
 \hline
 {\bf Workload} & {\bf Full Run} & {\bf Partial Run} & {\bf Error}\\
 \hline
 Blackscholes & 2 & 2 & 0\% \\ 
 Bodytrack & 3 & 2 & 33.33\% \\  
 Canneal & 1 & 1 & 0\% \\
 Ferret & 2 & 2 & 0\% \\
 Fluidanimate & 2 & 2 & 0\% \\
 Freqmine & 1 & 1 & 0\% \\
 Streamcluster & 3 & 3 & 0\% \\
 Swaptions & 3 & 3 & 0\% \\
 DGEMM & 5 & 6 & 20\% \\
 \hline
\end{tabular}
\captionsetup[table]{skip=6pt}
\caption{\it{CPU usage comparison of Static and Dynamic Profiling for well known application workloads. (Units: Number of cores)}}
\vspace{-1.5em}
\label{table:cpuaccuracy}
\end{table}
\section{Limitations of the proposed approach}
\begin{figure}[h!]
	\includegraphics[width=0.5\textwidth, height=4cm]{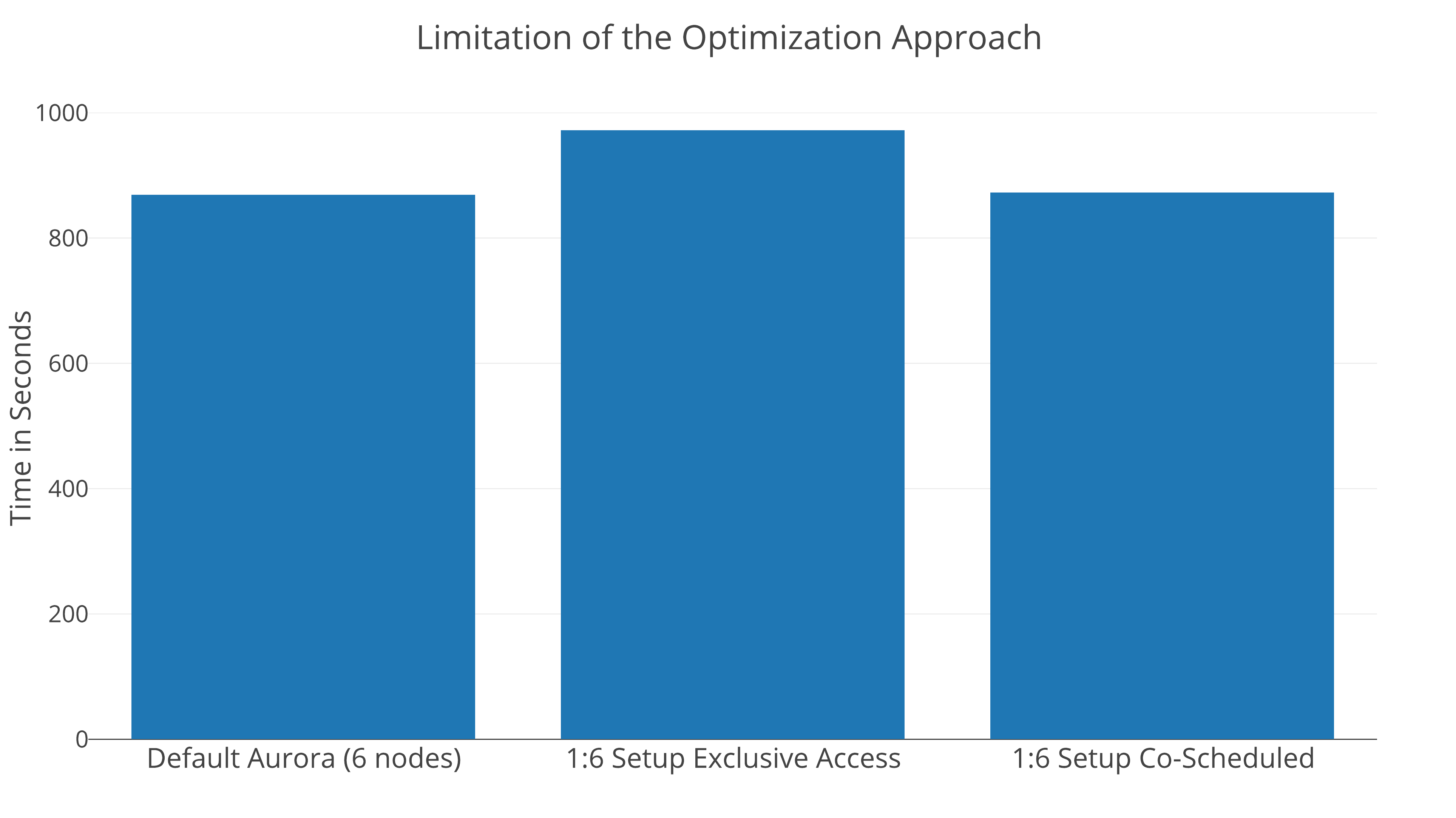}
	\caption{\it{Runtime comparison of different setups showing Limitation.}}
    \vspace{-1.5em}
    \label{fig:limitation}
\end{figure}

Our approach assumes that the resource requirements, for submitted applications, are over-estimated. If the user requests the right amount of resources, then the proposed system would add an overhead of the wait time for the application to go through the optimizer. Our system has a tradeoff between wait time and throughput, although we have an increased amount of wait time for each job, our overall throughput is much better. If the application comes with the right amount of resources, then expectedly, due to the optimizer overhead, there will be a negative impact on the throughput. 
In Figure \ref{fig:limitation}, we can observe that if the applications arrive with the right amount of resources then the Exclusive Access approach takes 103 seconds longer, while the Co-scheduled approach takes 4 seconds longer to finish. The Exclusive Access and Co-Scheduled approach also have an overhead of one VM/node, to host and execute the optimization module. 

\section{ANALYSIS}


We discuss the experimental results for the Exclusive Access and Co-scheduled Approach in terms of resource utilization and throughput.

\subsection{Default Aurora Results}
	The jobs in the default Aurora experiments had 50\% more resources allocated, than required, for memory and CPU. From our experiments, we found that the CPU utilization of the cluster was low -- at an average of 30-35\% for various setups. The same holds true for memory,  where the average utilization was 68-72\%. The graphs in Figures 7-12,  we compare both the approaches with an identical number of nodes for the default Aurora experiment.

\subsection{Exclusive Access Optimization Method}
In this Exclusive Access execution model, we pick one task at a time and run it in the little cluster to estimate its resources requirement. To estimate the optimum resource requirement we run a job for a few seconds and take periodic measurements to get the modulus of the standard deviation of the observation points. If the majority of the observations remain in a predefined confidence factor, then we calculate the resource requirements. Once we are done with the resource estimation of each job, we start adding it to the waiting queue of Apache Aurora. Aurora launches jobs to the cluster with our new estimated resource requirements.

\begin{figure}[h!]
	\includegraphics[width=0.5\textwidth, height=4cm]{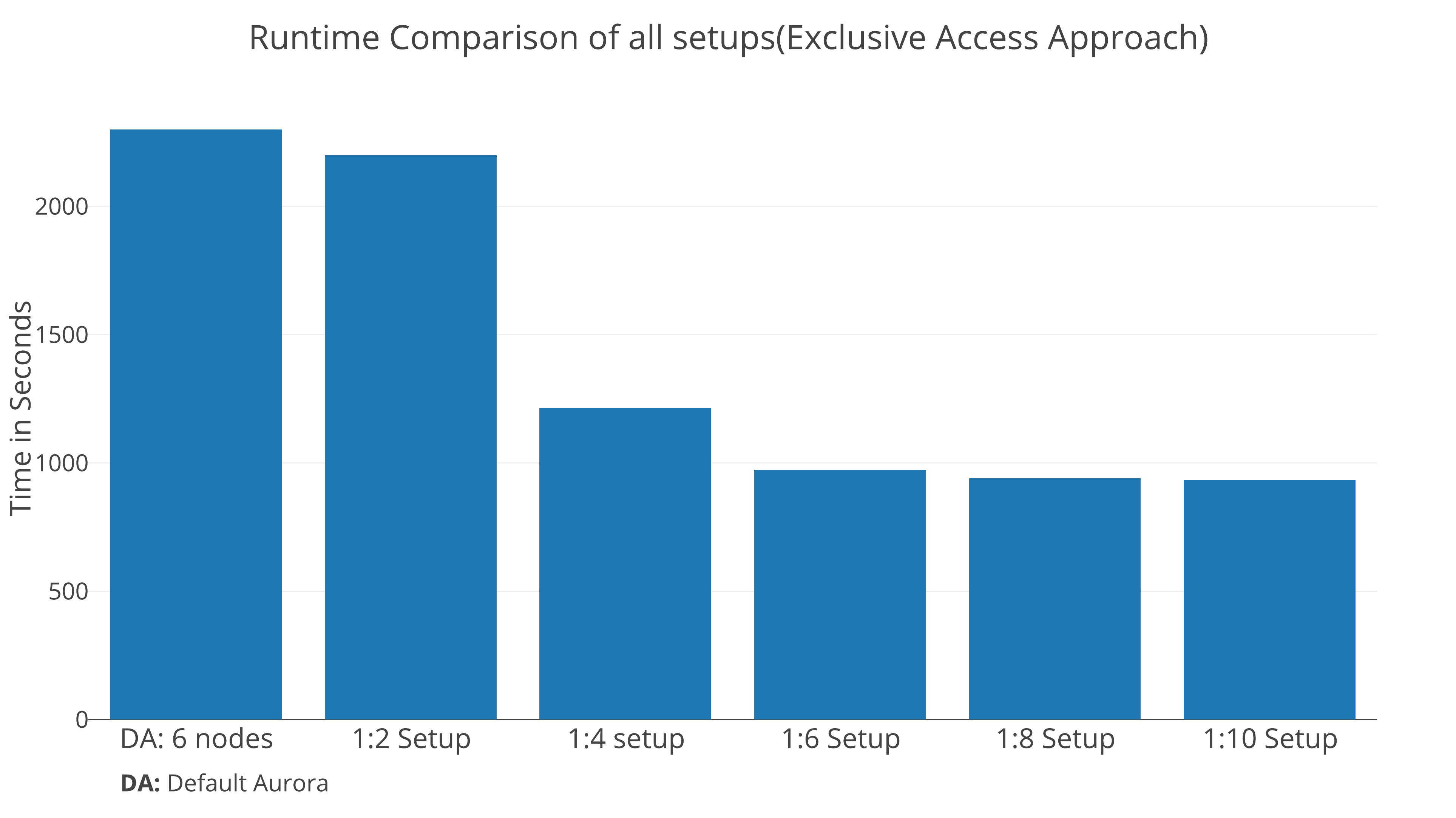}
	\caption{\it{Runtime comparison of different setups with Exclusive Access Optimization Method}}
    \vspace{-1.6em}
    \label{fig:queuedThrp}
\end{figure}

\begin{figure}[h!]
	\includegraphics[width=0.5\textwidth, height=4cm]{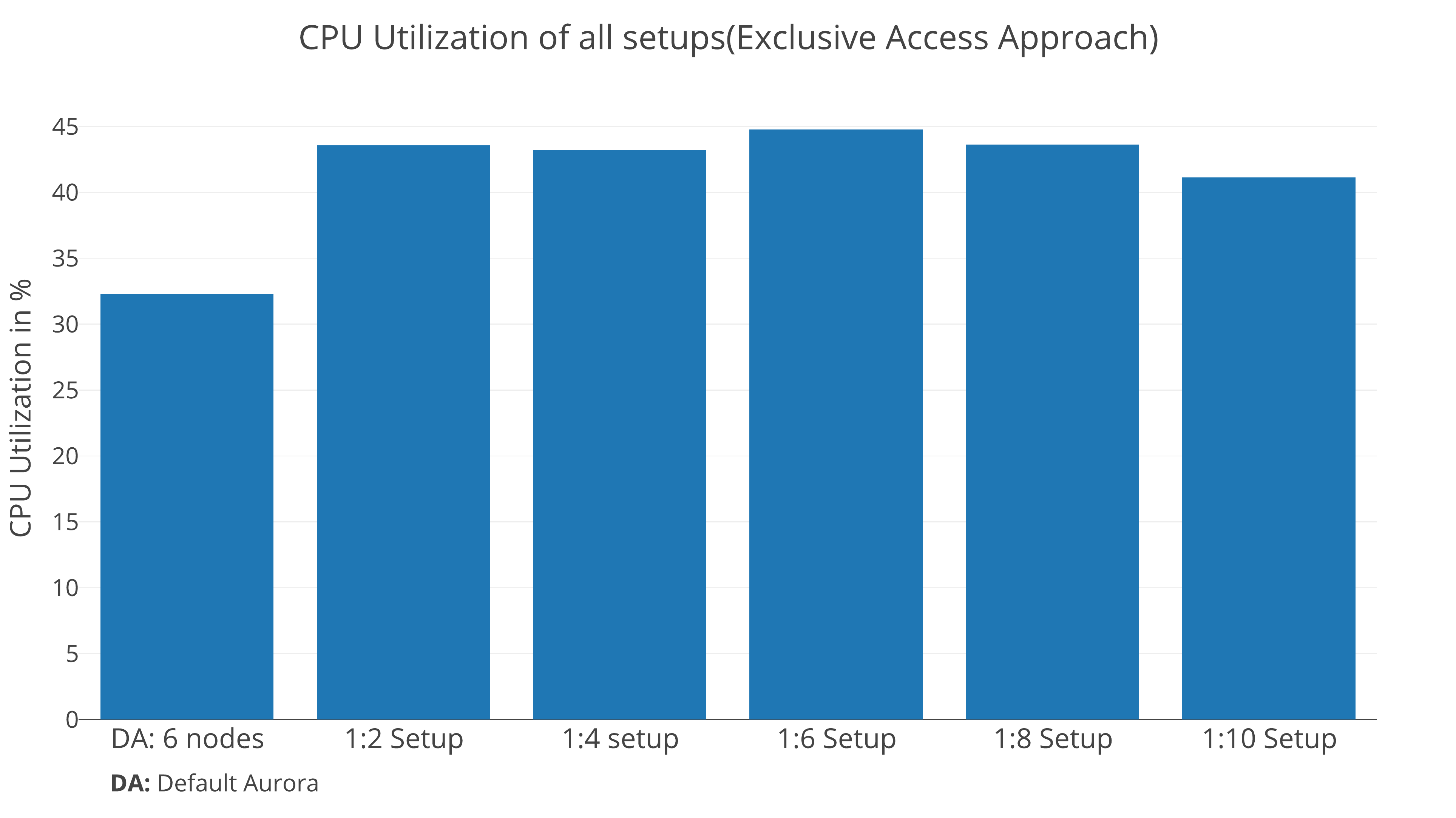}
	\caption{\it{CPU usage comparison of different setups with Exclusive Access Optimization Method}}
    \vspace{-1.6em}
    \label{fig:queuedCPU}
\end{figure}

\begin{figure}[h!]
	\includegraphics[width=0.5\textwidth, height=4cm]{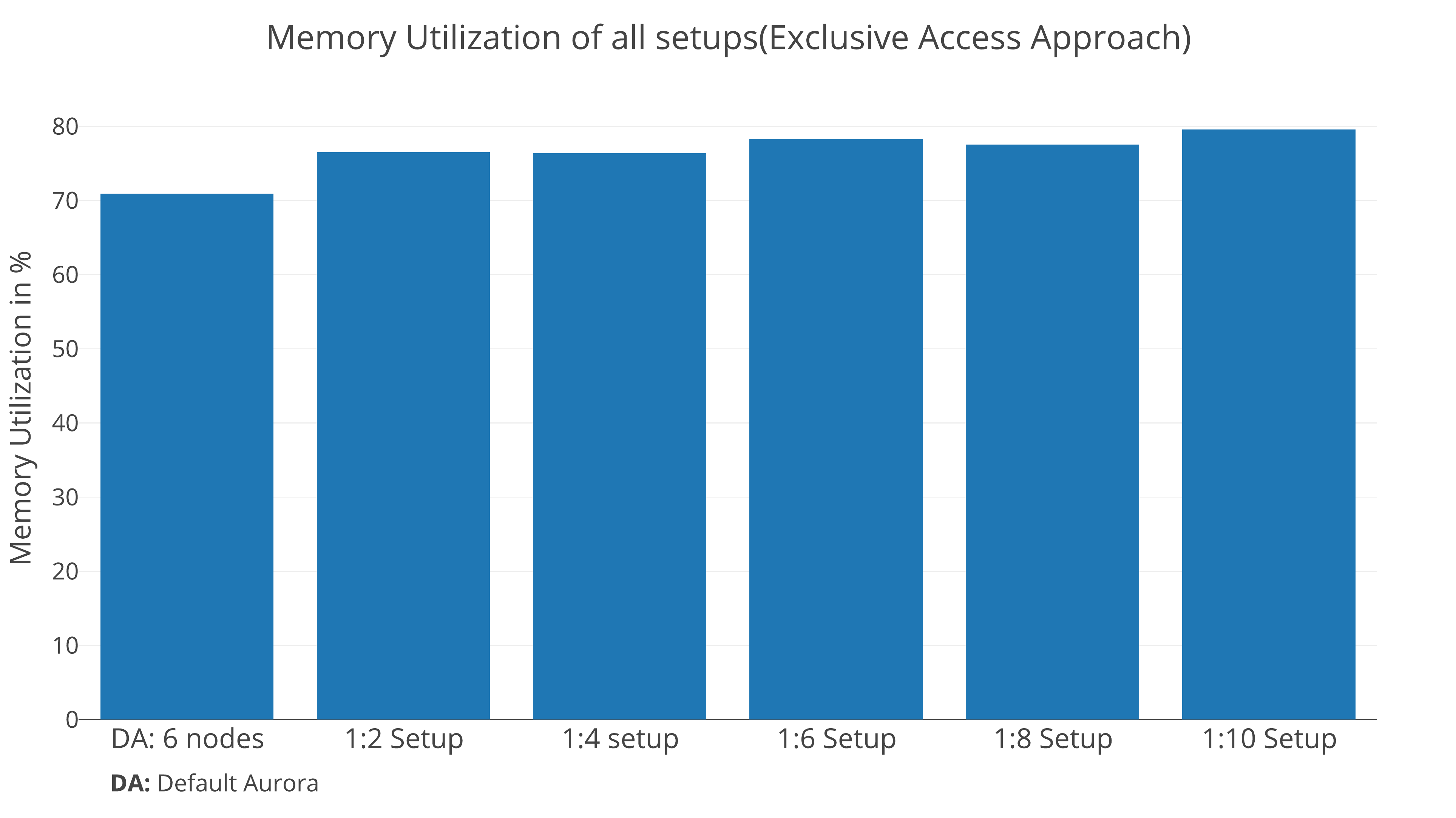}
	\caption{\it{Memory usage comparison of different setups with Exclusive Access Optimization Method}}
    \vspace{-1.6em}
    \label{fig:queuedMem}
\end{figure}

We ran the experiment with setups ranging from 1:2  to 1:10.  In Figure \ref{fig:queuedThrp}, we see that all the setups experience significantly better throughput. The time taken to launch and finish all the applications keeps decreasing and then stays constant for setups of 1:6 and 1:8. After this point, we see no improvement in throughput. Also, when we compare the throughput with the default Aurora setup with 6 nodes we find that there is an 81\% improvement with 1:6 setup. The reason why throughput does not decrease anymore is due to the optimization overhead. Each application gets exclusive access to the machine and this increases the time required to estimate the optimization. In terms of CPU utilization, referring to  Figure \ref{fig:queuedCPU}, the improvement is over 32\%. This is primarily due to the right sizing of the applications and the ability of Aurora to efficiently schedule the application, using First-Fit, on the nodes. In Figure \ref{fig:queuedMem}, the memory utilization across experiments seems to be very similar, but still, offers about 10\% improvement over the default Aurora setup. However, 1:10 achieves the highest Memory Utilization, and the 1:8 setup experiences a slight improvement in throughput when compared to the 1:6 setup. As these are marginal gains, we conclude that for the exclusive access optimization method it is best to have a ratio of 1:6 to 1:8.

\subsection{Co-Scheduled Optimization Method}
    In the Co-Scheduled execution model, we deploy multiple threads to launch multiple jobs in the little cluster to estimate their individual resource requirements in parallel. This parallel approach reduces the turn around time as multiple jobs can be submitted to the Aurora waiting queue, instead of one job at a time like in the Exclusive Access model. The process of calculating the optimal resource requirement remains the same as serial execution. 
 So the difference between serial and parallel approach is how quickly we can determine the optimum resource requirement of each job and submit it to the Aurora framework, which in turn co-schedules them via Mesos.
    
\begin{figure}[h!]
	\includegraphics[width=0.5\textwidth, height=4cm]{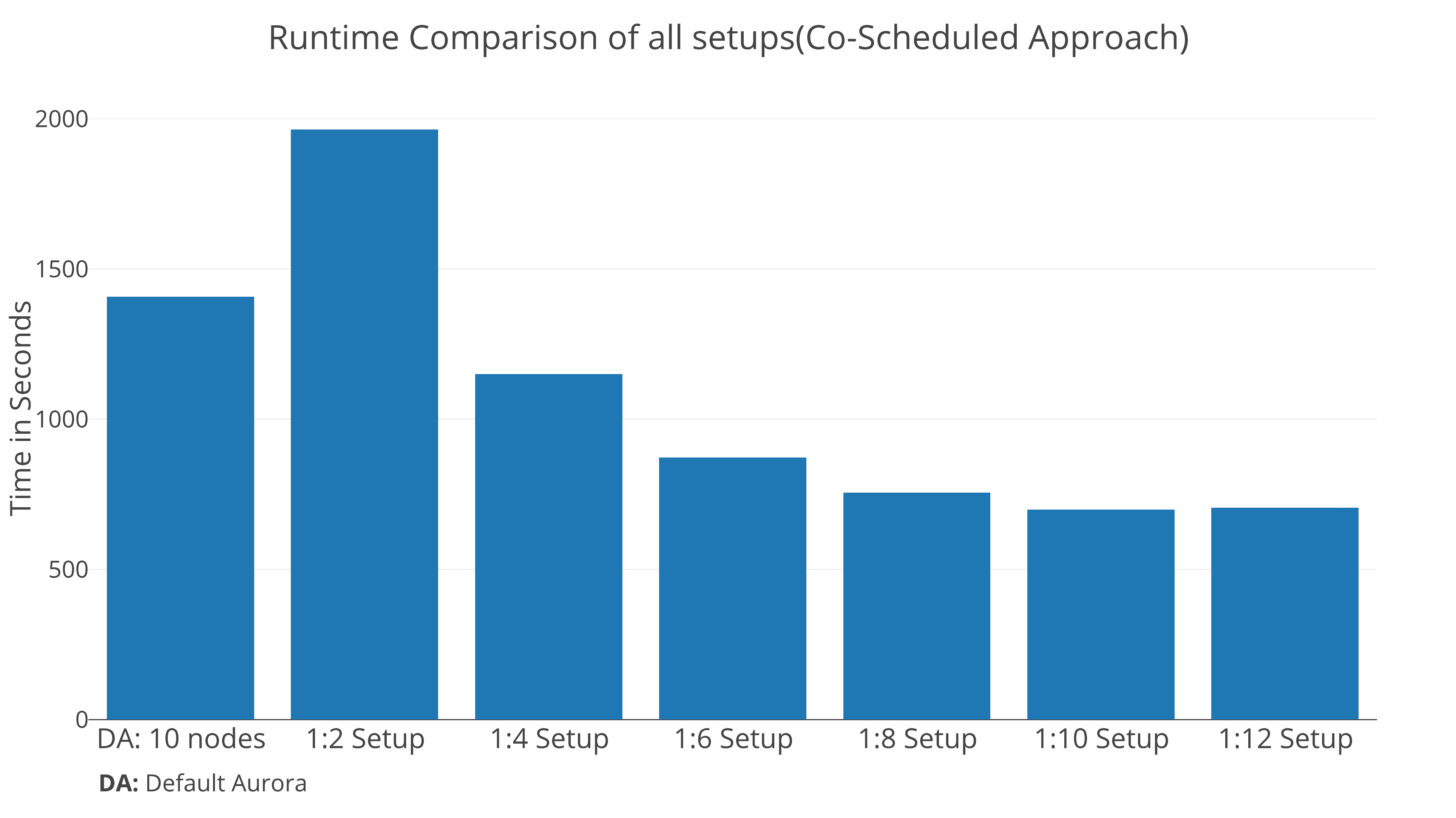}
	\caption{\it{Runtime comparison of different setups with Co-Scheduled Optimization Method}}
    \vspace{-1.5em}
    \label{fig:simultaneousThroughput}
\end{figure}

\begin{figure}[h!]
	\includegraphics[width=0.5\textwidth, height=4cm]{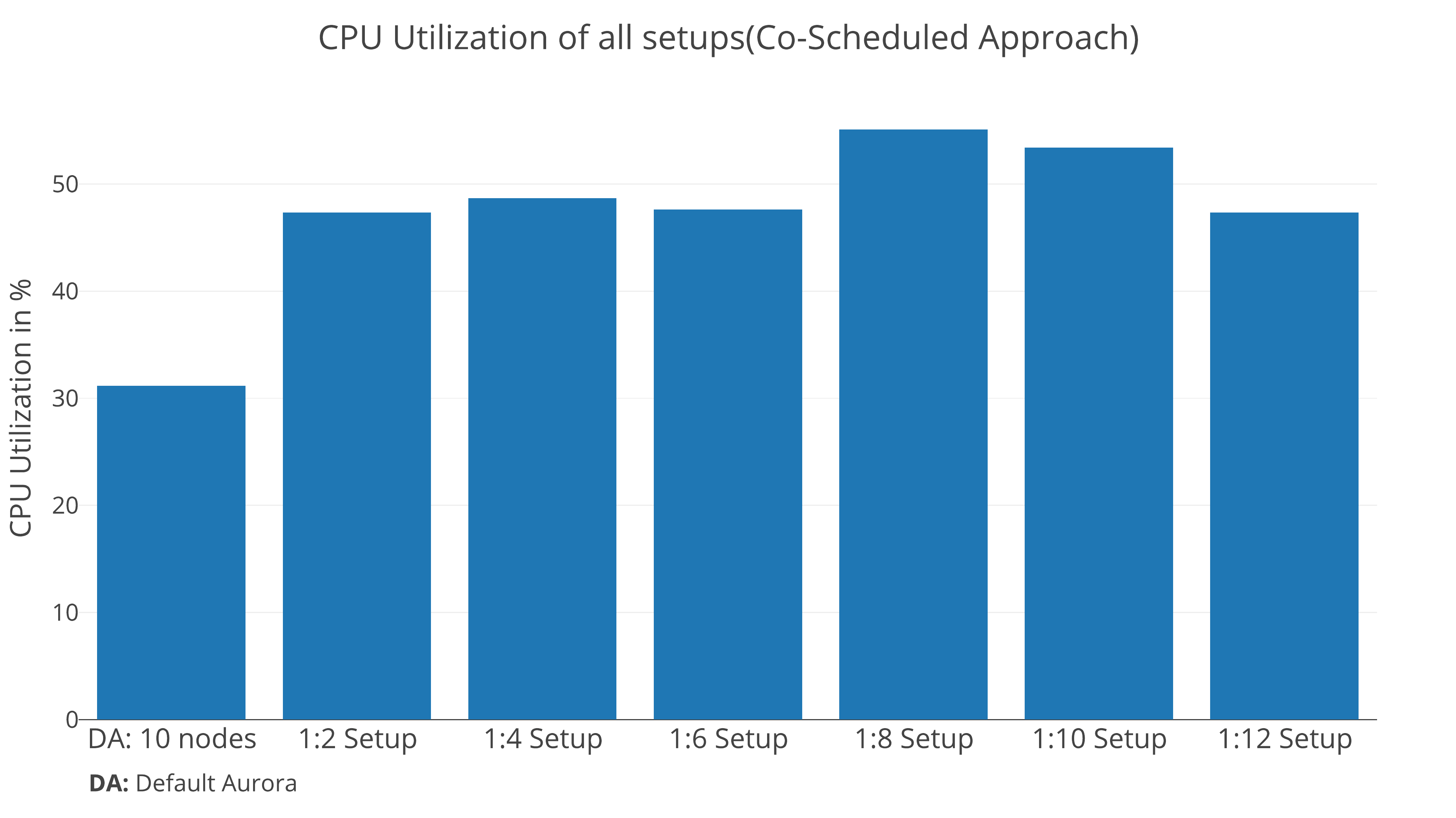}
	\caption{\it{CPU Usage comparison of different setups with Co-Scheduled Optimization Method}}
    \vspace{-1.5em}
    \label{fig:simultaneousCpu}
\end{figure}

\begin{figure}[h!]
	\includegraphics[width=0.5\textwidth, height=4cm]{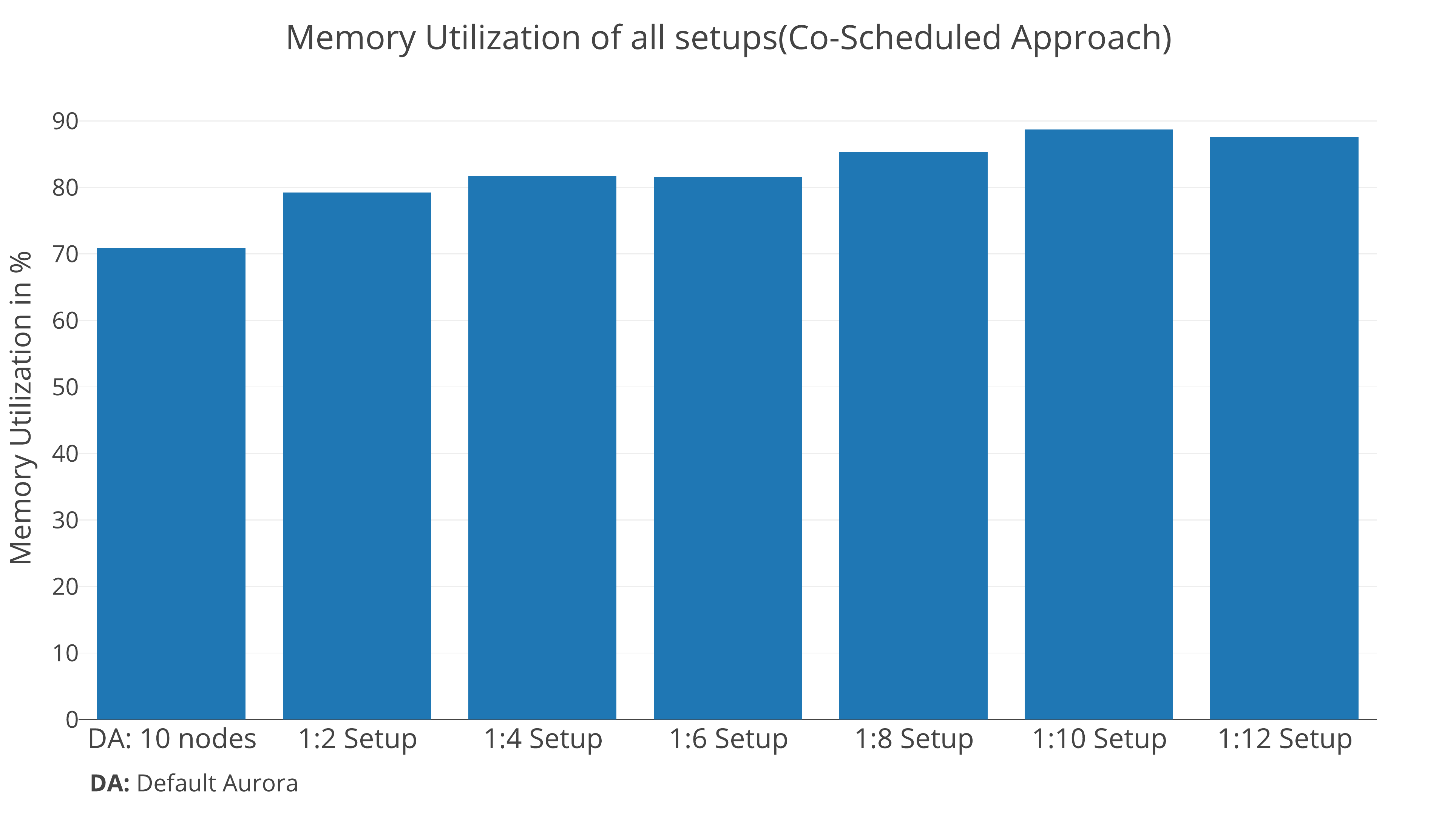}
	\caption{\it{Memory Usage comparison of different setups with Co-Scheduled Optimization Method}}
	\vspace{-1.5em}
	\label{fig:simultaneousMemory}
\end{figure}

As in the previous set of experiments with the Exclusive Access optimization method, we again ran experiments for different cluster setups ranging from 1:2 to 1:12. In Figure \ref{fig:simultaneousThroughput}, we see that the runtime has decreased by about 67\%. In Figure \ref{fig:simultaneousCpu}, we observe a significant improvement by 53\% in CPU utilization when compared with the default Aurora setup with 10 nodes. In terms of memory utilization, it can be seen in Figure \ref{fig:simultaneousMemory} that the utilization is highest with the 1:10 setup and improves by 22\% over the default Aurora setup with 10 nodes. By observing the data for throughput, CPU, and memory, we see that 1:10 setup performs the best, though 1:8 has a better CPU utilization than 1:10 by 3\%. The overall gains with 1:10 cluster size ratio are better with memory utilization by 4\% and throughput by about 8\%.
    
\subsection{Comparison}

\begin{figure}[h!]
	\includegraphics[width=0.5\textwidth, height=4cm]{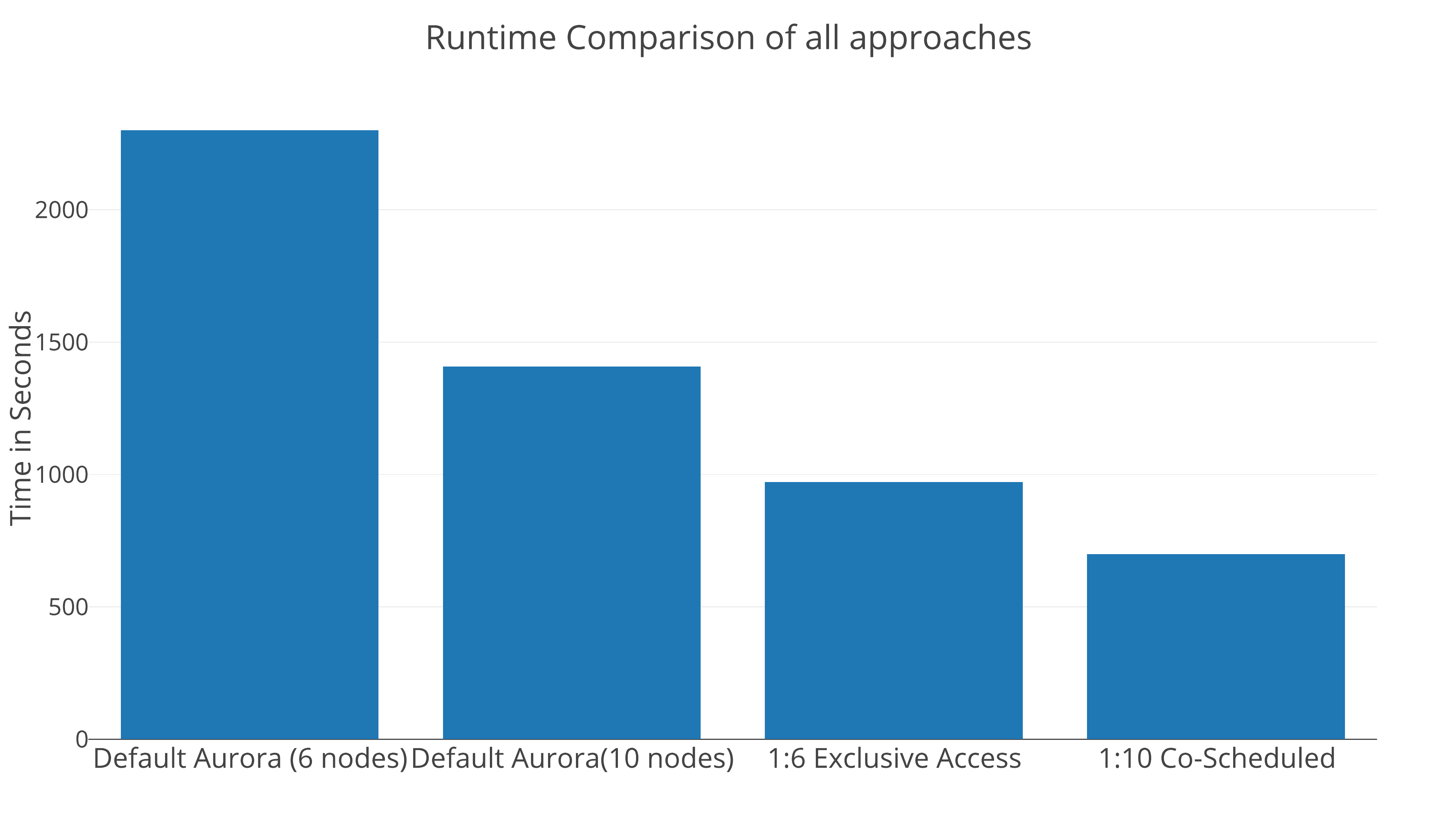}
	\caption{\it{Runtime comparison of different Approaches}}
    \vspace{-1.5em}
    \label{fig:compThr}
\end{figure}

Our results show that 1:6-1:8 setup is best for the Exclusive Access optimization method, and 1:10 is the best for the Co-Scheduled optimization method. We also compared the two approaches with each other. In Figure \ref{fig:compThr}, we see that the throughput of the Exclusive Access method is better than the default Aurora setup with 10 nodes by over 36\%. It is similar for CPU utilization, which is also much better at 35\%. There is also an improvement in the memory utilization by over 9\%. This shows that optimizing the resource requests can cut down the size of the cluster and still perform better than a bigger setup. With the Co-Scheduled approach, the time required to get an optimized estimate of the resources is much smaller when compared to the Exclusive Access approach. The Exclusive Access approach on an average takes about 450-500 seconds to optimize 90 applications, whereas the Co-Scheduled approach takes about 90-120 seconds to do the same. 

\begin{figure}[h!]
	\includegraphics[width=0.5\textwidth, height=4cm]{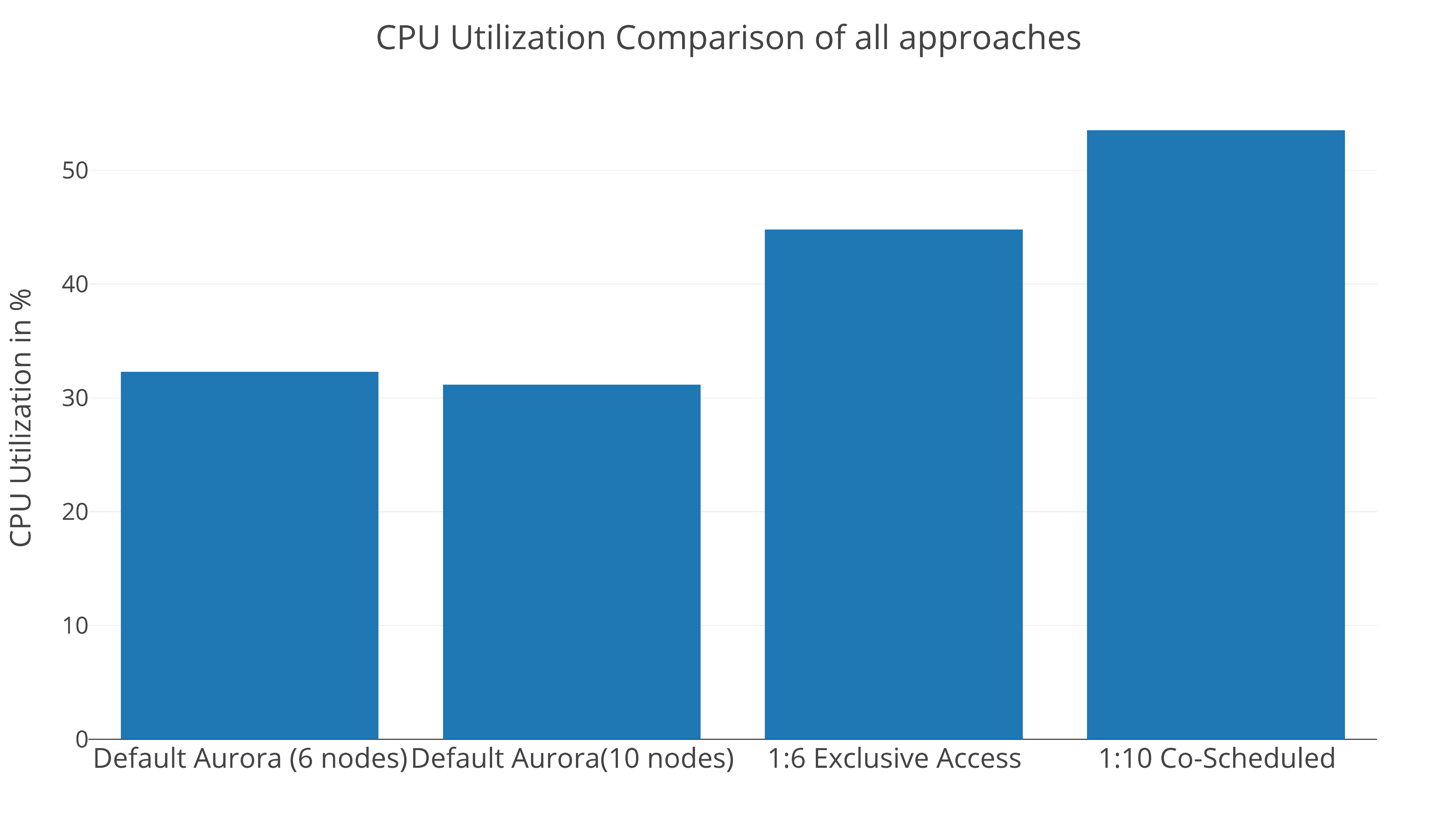}
	\caption{\it{CPU utilization comparison of different Approaches}}
    \vspace{-1.5em}
    \label{fig:compCpu}
\end{figure}

\begin{figure}[h!]
	\includegraphics[width=0.5\textwidth, height=4cm]{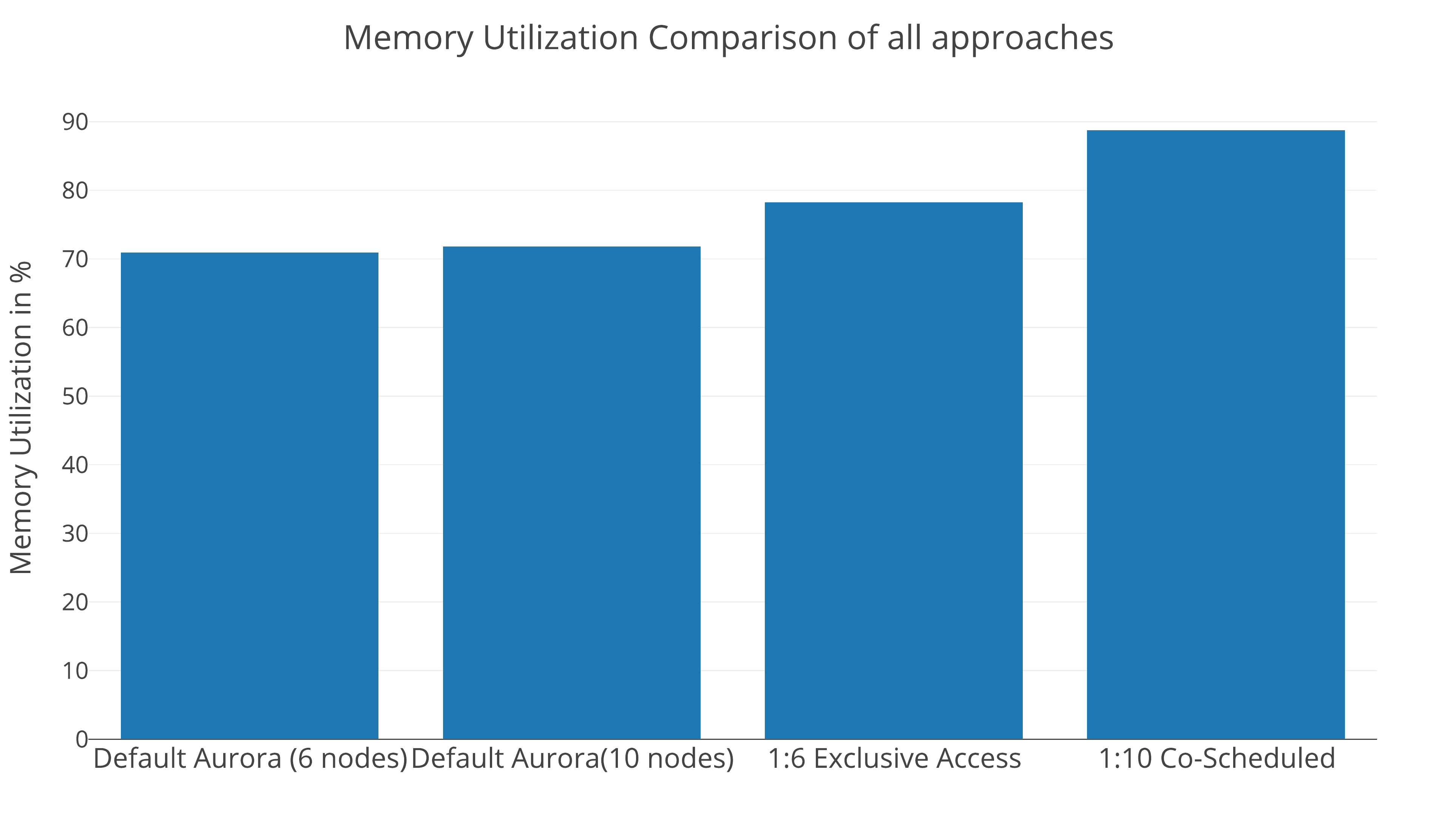}
	\caption{\it{Memory utilization comparison of different Approaches}}
    \vspace{-1.5em}
    \label{fig:compMem}
\end{figure}
\section{RELATED WORK}

A project on automating resource allocation in a Hadoop based system was conducted by Palden Lama and his team at the University of Colorado. Their system was named AROMA~\cite{Lama2012AROMA}, and it includes work related to optimize the resources allocated to a task. They use a regression model to analyze and predict a model for the task. It includes a set of models to choose from based on the regression analysis. Our work differs from this approach in that we analyze the resource utilization. AROMA requires previously collected information to base its decision and also groups similar tasks and provides same resource decision to all such tasks. Our approach is based on resource utilization of the current run with no previous data and also considers each task to be independent. Optimizing each task separately allows us to be more accurate with our optimizations.

In another study by Herodotos Herodotou et al., they experimented with a cost-based optimization approach to find the best configuration~\cite{HerodotouProfilingPrograms}. They propose a system with a profiler to collect (1) runtime information of a task, (2) a cost-based optimizer to find a good fit configuration, (3) and a what-if analyzer to check whether the suggested configuration is a good fit or not. They also used it to find if the suggested configuration would have any effects on the task. The proposed system is for map-reduce tasks on Hadoop. Our work differs in the way we run the task to find the optimal resources. Their system takes a decision on a sample of jobs of the same type whereas our approach considers all tasks independently.

Karthik Kambatla et al.~\cite{Kambatla2009TowardsCloud} conducted research on optimal provisioning of a Hadoop job by resource consumption statistics of the job. They demonstrated how resource requirement differs across different applications. They also discuss how unoptimized resource provisioning on the cloud increases costs as the user ends up paying for more resources. In our approach, we have statistically determined the resource requirement by profiling the job in a little cluster, with Docker API, before running the job on the big cluster for the full run.

A research project on Cost-Effective Resource Provisioning for MapReduce in a Cloud~\cite{Palanisamy2015Cost-EffectiveCloud} by Palanisamy et al. presents a resource management framework called {\it Cura} to discuss the cost-effectiveness of MapReduce for existing cloud services. Cura creates an automatic configuration for a job to optimize the resource allocation by leveraging the MapReduce profiling. Their target is to deal with the situation that consists of many short running jobs with lower latency requirements. 

Jiang Dejun et al.~\cite{Dejun2011ResourceClouds} conducted a study on resource provisioning for web applications in the cloud to show how to efficiently create a performance profile of each individual virtual machines instance. The performance profile of each machine help users to predict how adding a new virtual machine will lead to performance gain at a lower cost and what would be the best use of a newly acquired machine in any tier of a web application. Their research shows how to make accurate profiling of a machine efficiently in a heterogeneous cluster.

\section{Future Work}
In future, we will work on optimizing workloads whose resource usage varies significantly during the execution. We will also analyze the effect on performance when the application queue is altered. An area that also needs to be explored is optimizations for heterogeneous clusters. Also, a further improvement to the current system would be to include VM or container migration from the little cluster to the big cluster.
\section{CONCLUSION}

\begin{itemize}

\item The Exclusive access approach provides significant gains over the default way to use Mesos and Aurora. The results get even better when using the Co-Scheduled approach. Both the approaches  provide much better CPU and memory utilization too. Due to the improvements in the accuracy of the resource requests, Aurora and Mesos are able to better schedule applications on each node. This directly increases CPU and Memory utilization.

\item The optimizations using the exclusive access incur an overhead and so the throughput improvement is limited as the size of the big cluster increases. The Co-Scheduled approach is able to mitigate this problem. It optimizes the estimation of the resource requests much faster and we can see its effects in throughput.  
\item A limitation of the current approach is that an application has to be re-started on the big cluster. It is to be noted that Mesos is planning to provide support for VM migration, which will allow us to migrate applications from the little to the big cluster without a need to re-start. However, even with restarting the application, both the approaches preform much better than the default way with 81\% better throughput when comparing Exclusive access approach with default Aurora-Mesos setup with 6 nodes, 53\% in CPU utilization and 22\% in Memory Utilization when comparing Co-Scheduled approach with default Aurora-Mesos setup with 10 nodes. Our experimental setup allows the determination of the best ratio of machines on the little and big clusters for both approaches, i.e. 1:6-1:8 for Exclusive Access approach and 1:10 for the Co-Scheduled approach. 

\end{itemize}

\addtolength{\textheight}{-12cm}   





\bibliographystyle{IEEEtran}
\bibliography{IEEEabrv,Mendeley}{}
\end{document}